# On the Role of Low Energy Modes of the Flow on Sub-Grid Scale Parameter Prediction

Hossein Rahmani,[1,a)] Hamid Kalaei,[1] Ghasem Akbari,[2,b)] , Nader Montazerin,[1,c)]

[1] *Department of mechanical engineering, Amirkabir University of Technology, Tehran, Iran.*

[2] *Faculty of industrial and mechanical engineering, Qazvin branch, Islamic Azad University, Qazvin, Iran.*

[a)] *Electronic mail: rhmnhossein@gmail.com, Tel: +98 936 609 5744*

*LinkedIn account: https://www.linkedin.com/in/hossein-rahmani-325943b7/.*

[b)] *Electronic mail: g.akbari@qiau.ac.ir, Tel: +98 912 204 4399*

*LinkedIn account: https://www.linkedin.com/in/ghasem-akbari-72241443/.*

[c)] **Corresponding Author**. *Electronic mail: mntzrn@aut.ac.ir, Tel: +98 912 159 3541*

*LinkedIn account: https://www.linkedin.com/in/nader-montazerin-336308102/.*

# On the Role of Low Energy Modes of the Flow on Sub-Grid Scale Parameter Prediction

Abstract:

Large Eddy Simulation is based on decomposition of turbulent flow structures to large energy containing scales and small subgrid scales. The present study captures all flow low energy modes of a sample shear layer using the proper orthogonal decomposition (POD) method. In order to analyze the role of flow low energy modes on subgrid scale parameter prediction, the a-priori approach is chosen on a stereoscopic particle image velocimetry (SPIV) data that its missing/erroneous data is reconstructed using the gappy POD method. Particularly, similarity and mixed models are used to evaluate the SGS parameters. The results of the mixed model are compared, before and after removing the high energy modes. It is shown that SGS stress is predicted more accurately in the mixed model after removing the highest energy mode.

## 1. Introduction

Large eddy simulation (LES) simulates larger three-dimensional unsteady turbulent structures directly and also models the effect of smaller-scale flow structures. LES can be more accurate and reliable than Reynolds averaged Navier Stokes (RANS) approach for flows in which large-scale unsteadiness is significant. The vast computational cost of explicit representation of the small-scale motions is also avoided in LES as compared with direct numerical simulation (DNS) [1].

The filtered Navier-Stokes equation is defined as:

$$\frac{\overline{D}\,\overline{U}_j}{\overline{D}t} = \nu \frac{\partial^2 \overline{U}_j}{\partial x_i \partial x_i} - \frac{\partial \tau_{ij}}{\partial x_i} - \frac{1}{\rho}\frac{\partial \overline{p}}{\partial x_j}, \qquad (1)$$

where $\overline{U}_j$ represents the filtered velocity components and $x_j$ denotes orthogonal coordinates ($j = 1, 2, 3$ for $x, y$ and $z$ directions, respectively) and $\frac{\overline{D}}{\overline{D}t} = \frac{\partial}{\partial t} + \overline{\mathbf{U}}\cdot\nabla$ is the material derivative operator. Additionally, $\tau_{ij}$ notifies SGS stress components which should be modeled and are defined as:

$$\tau_{ij} = \overline{U_i U_j} - \overline{U}_i\,\overline{U}_j. \qquad (2)$$

## 1.1 SGS models

The Smagorinsky, similarity and mixed models are three well-known SGS models that are introduced and compared as follows.

### 1.1.1 Smagorinsky model

The most common eddy-viscosity subgrid-scale model of LES is the Smagorinsky model [2], which is presented as

$$\tau_{ij} = -2(C_{Smag}\Delta)^2 |\overline{S}|\,\overline{S}_{ij} \qquad (3)$$

where $C_{Smag}$ is the Smagorinsky coefficient, $\Delta$ is the filter width, $\overline{S}_{ij}$ is the filtered strain rate, and $|\overline{S}|$ is the filtered strain rate magnitude ($|\overline{S}| = \sqrt{2\overline{S}_{ij}\,\overline{S}_{ij}}$). This model is simple and stable and is therefore frequently used. It only predicts forward scatter of energy and produces little correlation with the real stress $\tau_{ij}$ [3, 4].

*1.1.2 Similarity model*

Bardina *et al.* [5] postulated the similarity model based on the similarity between the velocity structures at scales larger and smaller than the filter scale. According to the similarity model, SGS stress tensor is approximated as

$$t_{ij} = C_{Sim}\left(\overline{\overline{\overline{U_i}\,\overline{U_j}}} - \overline{\overline{\overline{U_i}}}\,\overline{\overline{\overline{U_j}}}\right) \tag{4}$$

where $C_{Sim}$ is the similarity model coefficient, the first filter is denoted by the lower overbar and the second filter (test filter) is shown by the upper overbar. The length scales of the first and second filters are the same in the original Bardina model, however, in a modified and more accurate version, the size of the second filter is considered larger than the size of the initial filter [3]. This model and its variants exhibit considerable correlation with the real stress [3, 5] and additionally predict back-scatter of energy.

*1.1.3 Mixed model*

The similarity model under-predicts the SGS dissipation of energy which intensifies flow fluctuations and makes LES unstable [6]. An eddy-viscosity term (especially based on the Smagorinsky model) is added to the similarity model in order to overcome this problem, correct the prediction of the SGS dissipation of energy and make the model more stable. This is known as the mixed model which is formulated as:

$$t_{ij} = C_{Sim}\left(\overline{\overline{\overline{U_i}\,\overline{U_j}}} - \overline{\overline{\overline{U_i}}}\,\overline{\overline{\overline{U_j}}}\right) - 2(C_{Smag}\Delta)^2 \left|\overline{S}\right|\overline{S}_{ij} \tag{5}$$

*1.2 Multiscale approach in SGS parameter modeling*

The idea of using multiscale approach to solve the Navier-Stokes equation was initially used by Temam and his colleagues [7]. Based on this method, they decomposed the velocity into small and large eddies and consequently, decomposed the Navier-Stokes equation into two coupled equations. The multiscale approach for turbulence modeling was espoused by Hughes *et al*. [8], where they only applied the Smagorinsky model into the equation for the small eddies. They asserted that many shortcomings of the Smagorinsky based models originate from their inability to isolate between large and small flow scales. They showed that even with a constant-coefficient Smagorinsky model in the small scale equation, many shortcomings of the traditional LES models are obviated. Couplet *et al.* [9] assessed the intermodal energy transfer in a proper orthogonal decomposition of a turbulent separated flow. They stated that the Fourier-decomposition-based concepts of forward and backward energy cascades are also valid in the POD basis in which a net forward energy cascade exists. They noted the similarities between Fourier decomposition and POD and concluded that small POD modes (energetic modes) are associated with the large vortical structures and the large POD modes (low energy modes) are associated with small vortical structures.

In present study, the multiscale approach inspires evaluation of the role of low energy modes on SGS parameter prediction. These modes are expected to represent small vortical structures [9]. The Navier-Stokes equation remains tactless and low energy modes are directly used to calculate SGS model parameters. Herein, the a-priori approach is used to conduct such analysis and the SGS parameters are predicted using the similarity and mixed models.

## 1.3 A-priori evaluation of SGS models

One method for evaluating a SGS model is to simulate a particular flow and compare the results with experimental data, which is known as a-posteriori analysis [3]. In another method, named a-priori analysis, the fully resolved velocity fields obtained by DNS or experimental measurement are used to compare the local instantaneous SGS stresses with those predicted by SGS models. A-priori analysis based on the DNS data is restricted to low Reynolds numbers which is a fundamental drawback, however, experimental data provides access to high Reynolds numbers and more complicated flows.

Vreman *et al.* [10] conducted a-priori tests of large eddy simulation of the compressible plane mixing layer using DNS data of two-dimensional (2D) and three-dimensional (3D) flows. They particularly studied the magnitude of all sub-grid terms, the role of discretization errors, and the correlation of turbulent stress tensor with those predicted by Smagorinsky, Clark, and Bardina's scale similarity models. They reported that the Smagorinsky model, which is purely dissipative, shows poor correlations, however, the other two SGS models correlate considerably better [10]. A-priori analysis of some SGS models, using the DNS data of homogeneous isotropic turbulence, was performed by Martin *et al.* [11]. The authors argued that the scale similar models provide more accurate prediction of SGS stresses and heat fluxes in comparison with eddy viscosity and eddy diffusivity models. Besides, they showed that the scale similar models predict the SGS turbulent diffusion, SGS viscous dissipation, and SGS viscous diffusion more accurate than eddy viscosity and eddy diffusivity models.

A-priori analysis of a generalized Smagorinsky model, a stress similarity model, and a gradient model was performed, using the DNS data of isotropic turbulence decay in a compressible flow, by Pruett and Adams [4]. They compared the exact and modeled SGS stresses, component by component, based on correlation coefficients and stated that the Smagorinsky model correlates poorly against exact stresses (less than 0.2), gradient model correlates moderately well (approximately equal to 0.6), and the similarity model correlates considerably well (0.8 to 1). Okong'o and Bellan [12] evaluated performance of the constant-coefficient Smagorinsky, gradient, and scale similarity models based on an a-priori approach. They used the DNS data of a 3D temporal mixing layer with evaporating drops. The results show that the gradient and scale similarity model provide excellent correlation with the SGS quantities while the Smagorinsky model doesn't show good correlation. In another study, Okong'o and Bellan [13] assessed the performance of Smagorinsky, gradient, and similarity models for large eddy simulations of supercritical binary-species mixing layers. They utilized the DNS data of aforementioned flow to show that the Smagorinsky SGS fluxes again correlate poorly with exact quantities and the gradient and similarity models provide high correlated SGS fluxes. Akbari and Montazerin [14], applied a-priori approach and used experimental data to evaluate the performance of Smagorinsky and similarity models for the complicated flow field of a centrifugal turbomachine. The authors studied the correlation coefficient between the exact and modeled SGS quantities and the functionality between SGS stress/dissipation and resolved flow parameters. After calculating the joint probability density function between exact and modeled SGS stress/dissipation, the authors showed that the similarity model, which is capable of capturing energy backscatter, provides considerably larger correlations in comparison with Smagorinsky model. The functionality between the SGS stress/dissipation predicted by Smagorinsky model and resolved strain rate tensor, particularly in the case of cross components, was too weak.

All reviewed literatures report low correlation between the Smagorinsky and exact SGS quantities. Many of them in recent years tried to modify the Smagorinsky model and increase its accuracy. Among various modified versions of Smagorinsky model, the dynamic version introduced by Germano *et al.* [15], provides more exact predictions compared with the original model. Smagorinsky based models still have deficiencies [6, 16] which Hughes *et al.* [8] attribute them to their inability to successfully isolate the contribution of large and small scales of the flow.

Among many other studies that conducted a-priori analysis of SGS models, some used a-priori analysis as a tool to improve performance of the SGS models. Those works studied SGS parameters such as viscous/scalar dissipation [17]; divergence of the SGS stresses, and the kinetic energy transfer term [18]; and orientation and magnitude of SGS scalar flux [19]. In addition, they evaluated some SGS models including viscous/scalar dissipation closures [17]; scale similarity, one-equation viscosity and non-viscosity dynamic structure models [18]; and Smagorinsky, Vreman and Gradient models [19]. Particularly, Vaghefi et al. [17] proposed a modification in the SGS viscous dissipation model to improve its predictions, based on a-priori analysis.

*1.4 Present work*

In the present study, the SPIV velocity data of a turbulent mixing layer flow is used. A-priori approach is considered as a tool to evaluate the role of low energy modes of the flow in the prediction of SGS parameters. In order to conduct such analysis, the similarity and mixed models are selected to predict the SGS parameters. The purpose is to isolate the contribution of low energy modes of the flow on the prediction of SGS parameters. The POD method is used to capture different modes of the flow that are sorted based on their contribution on total kinetic energy. Thus, the Smagorinsky term of the mixed model is considered and the strain rates in this term are calculated before and after removing the high energy modes, to investigate the changes appeared in the SGS parameter predictions made by the mixed model.

## 2. Experimental data

### 2.1 Experimental setup

The SPIV data of the present study is taken from a turbulent mixing layer between two uniform, parallel streams of different velocities. A horizontal splitter plane separates these streams in a wind tunnel (Figure 1). The lower stream goes through a honey comb structure which provides a smaller velocity as compared with that of the upper stream. The honeycomb is constructed from an array of circular plastic tubes, each with a length of 0.2 m and a diameter of 14 mm. The splitter tip is sharp with a 12° tip angle. The test section is 2 meters long with a 0.3 m×0.3 m cross section. The inlet nozzle provides a uniform flow in the test section. An outlet transition minimizes the effects of the centrifugal fan on the flow in the test section. A circular gap before the fan inlet controls the required flow rate through the tunnel.

Three instantaneous velocity components are measured in a central vertical plane inside the shear layer using SPIV. The SPIV setup consists of the following components [14, 20]:

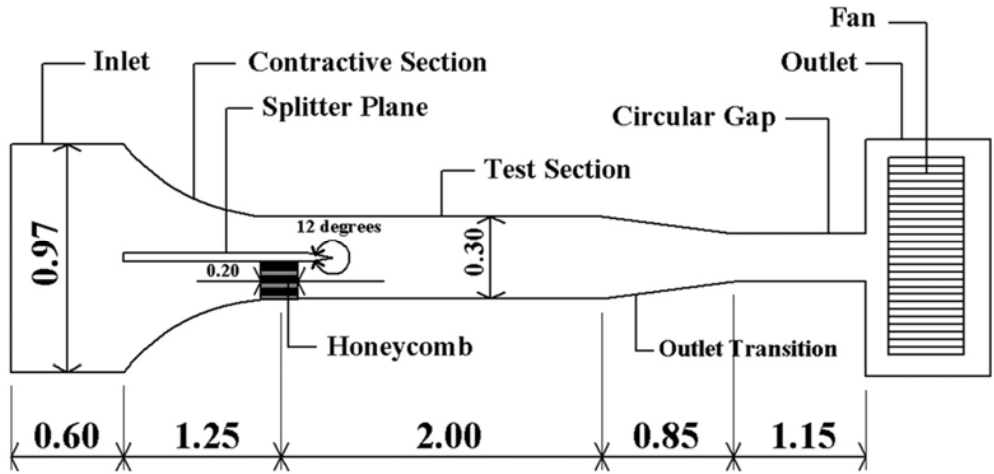

Figure 1. Schematic of side view and dimensions (in meters) of wind tunnel components.

- A double-cavity Quantel Brilliant Nd-YAG laser with 532 nm wavelength pulse and 150 mJ/pulse energy; the laser pulse frequency is 10 Hz.
- An optical guide system that delivers laser beams to the test section and converts the circular beam into a light sheet with adjustable thickness.
- Dantec FlowMap system hub that synchronizes laser pulses with camera aperture action.
- SAFEX F2010$^{pluse}$ fog generator that generates seed particles with the mean diameter of $1 \mathrm{m}m$ from evaporating the SAFEX standard fog fluid.
- Two FlowSense 1600´1186 pixel double-frame CCD cameras, equipped with AF Micro Nikkor lenses with 60 mm focal length and 532 nm interferential filters.

The schematic of measurement setup is presented in Figure 2. The angle of cameras with respect to normal direction of the measurement plane is $30°$.

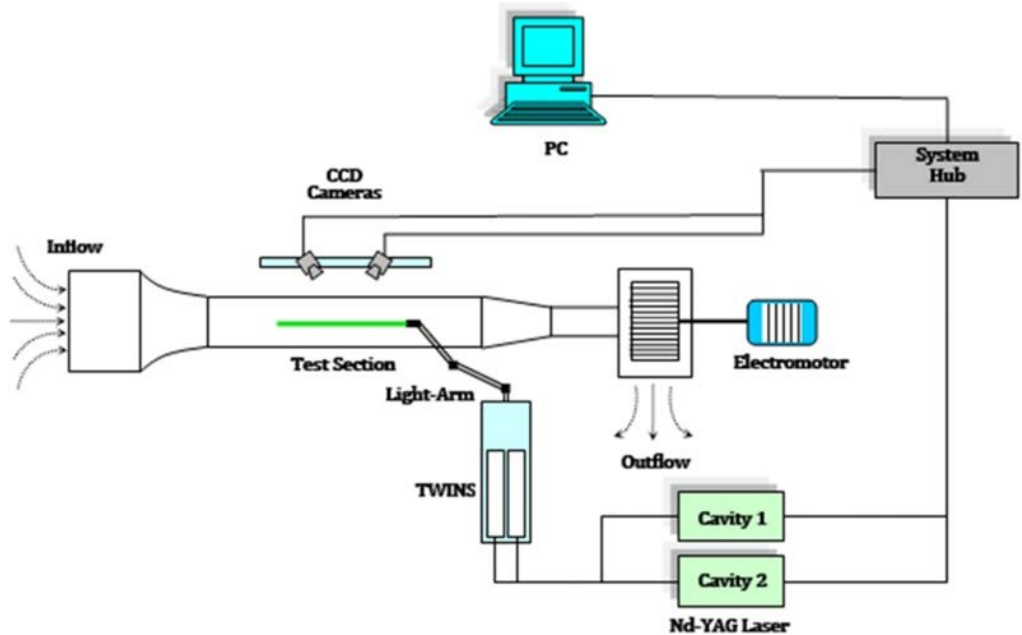

Figure 2. Schematic of experimental setup.

*2.2 Data acquisition*

Figure 3 shows the coordinate system and the field of view (FOV) position. The centers of the two selected FOVs are located at *x*=50 cm and *x*=90 cm (the origin of coordinate axes is at the tip of splitter plate). Self-similar results were only obtained for the measurements at *x*=90 cm. Therefore the data at *x*=90 cm are used in the present study. The FOV size at *x*=90 cm, is 15.2 cm ´ 10 cm which completely covers the mixing layer boundaries. The FOV grid points are 1.75 mm ~ 13$\eta$ and 1.47 mm ~ 11$\eta$ apart, in *x* and *y* directions, respectively, in which $\eta$ is the Kolmogorov length scale. The detailed information about the mixing layer flow of the present study is given in Table 1.

The temporal period between two successive image frames captured by each camera is set to 20 $\mu s$ and 1050 double frame images are captured for each FOV by each of two cameras. Adaptive cross-correlation is performed on 64´64 pixel interrogation areas to obtain 2D velocity vector maps. Each pair of these 2D maps (corresponding to two cameras) is combined to obtain a three-component vector map for each snapshot.

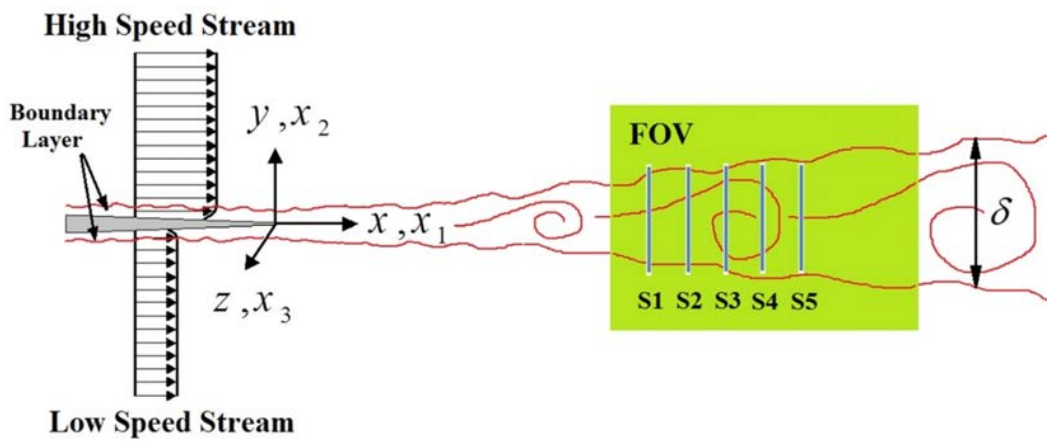

Figure 3. Schematic of mixing layer flow, the coordinate system and the position of FOV; (S1-S5 are the locations where self-similarity of velocity data is examined, as presented in Figure 6).

Table 1. Detailed information of mixing layer flow.

| Quantity | Value |
| --- | --- |
| High and low speed streams velocities | $U_h$ = 24.36 m/s and $U_l$ = 14.75 m/s |
| Mean velocity, $U_{mean}$ | 19.55 m/s |
| Integral length scale, $l_0$ | 0.1 m |
| Average of momentum thickness at the measurement plane, $\theta$ | 0.0089 m |
| Taylor length scale, $l_T$ | 0.0047 m |
| Turbulence Reynolds number, $Re_L = u' l_0 / \nu$ where $u'$ is the fluctuating velocity | 6863.00 |
| Reynolds number based on $\theta$, $Re_\theta = U_{mean} \theta / \nu$ | 11548.82 |
| Taylor Reynolds number, $Re_l = u' l_T / \nu$ | 320.85 |

## 2.3 Velocity field measured by SPIV

The SPIV is utilized to measure the velocity components $U_1$, $U_2$ and $U_3$ in $x$, $y$ and $z$ directions, respectively. $x$ and $y$ directions are in SPIV measurement plane, while the $z$ direction corresponds to the normal out-of-plane direction (Figure 3). The SPIV uncertainties originate from random and bias errors. The random error results from features like light noise which are not predictable, but should be minimized. The bias error depends on the systematic features of the experimental setup. The bias error due to particle tracking lag of the present SPIV measurements is evaluated to be 0.015% [21]. The displacement estimation error is another source of bias error that is evaluated to be about 0.62% for a 64×64 pixel interrogation area with Gaussian sub-pixel interpolation. The third important bias error is due to angular configuration of the cameras in SPIV setup and determines significance of out-of-plane velocity uncertainty in comparison with the in-plane velocity uncertainties. Accordingly, the out-of-plane to in-plane error ratio is about 1.8 when a symmetric configuration is used for the cameras with a 60 degrees angle between them [21]. The overall bias error in the worst condition is the sum of different mentioned sources, which is about 0.63% for each of the in-plane velocity components, about 1.13% for the out-of-plane velocity component, and about 1.44% for the velocity magnitude.

In order to conduct the a-priori analysis, the complete set of experimental data is required. However, the present SPIV data contains erroneous vectors which need to be reconstructed. Therefore, prior to the a-priori analysis, the gappy proper orthogonal decomposition (GPOD) method, as an efficient tool, is used to reconstruct and fill the missing data.

The ensemble average of planar velocity components in the mixing region as well as the vorticity component in z direction are calculated using 800 snapshots and are presented in Figure 4. The span-wise velocity ($U_2$), which provides mixing, is negative in the mixing region with the largest absolute values in the center of FOV. This means that the high speed flow moves towards the low speed layer [22]. The maximum out-of-plane vorticity component is in the mixing region and is contributed by shear stress growth and the added momentum exchange between high and low speed flows.

Figure 5 examines convergence of stream-wise velocity and the out-of-plane vorticity components at six selected points (as shown in Figure 4(a)). This figure shows that an accurate convergence of cumulative averages for stream-wise velocity could be provided even after the first 200 snapshots and only minor changes occur after 400 snapshots. Additionally, for other velocity components, the results were similar to those of stream-wise component. However, the suitable convergence for the out-of-plane vorticity component is after 400 snapshots.

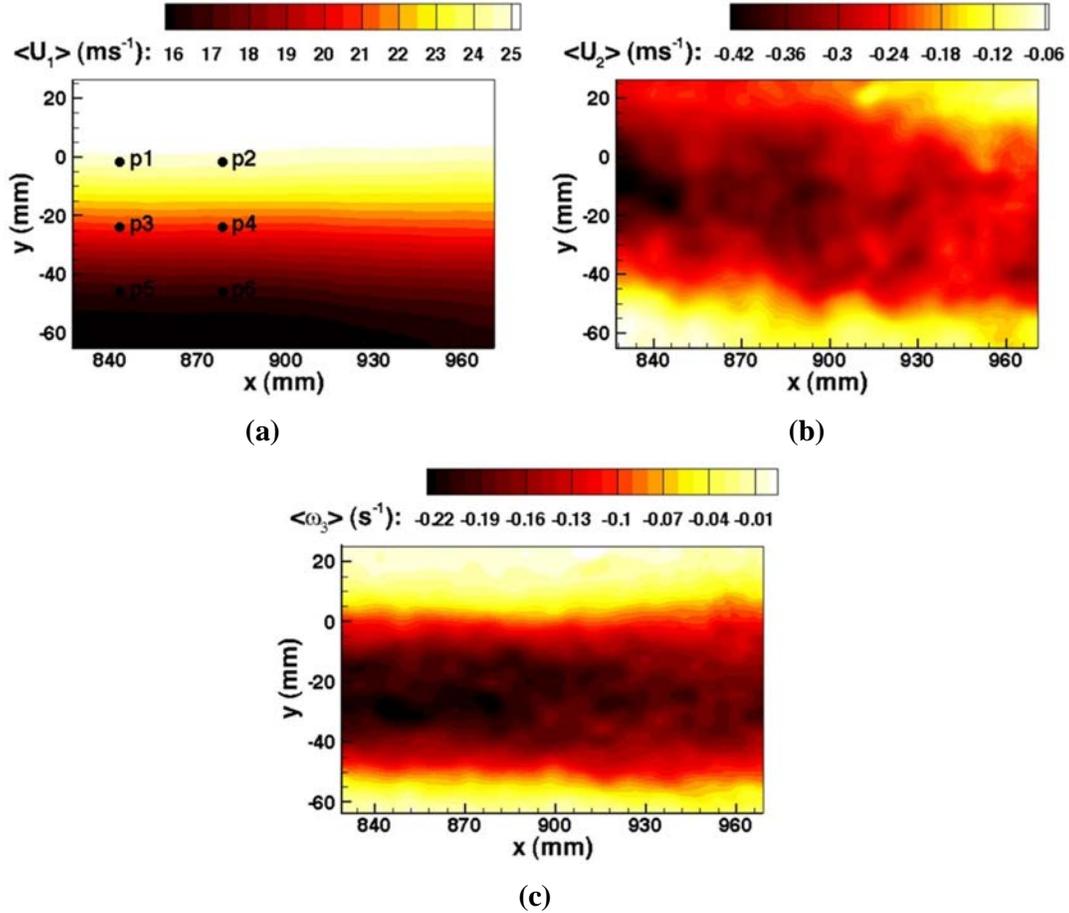

Figure 4. Ensemble average of two velocity components and the out-of-plane vorticity component; (a) velocity component in x direction, (b) velocity component in y direction, (c) vorticity in z direction. The location of six selected points for convergence assessment is presented in (a).

Figure 6 presents variations of ensemble average of normalized total velocity for five different sections in x direction (Figure 3), against normalized y axis. $U$ represents the total velocity, $\Delta U$ is the velocity difference between the total velocities of streams out of mixing layer ($\Delta U = U_2 - U_1$) and $\delta$ and $y_0$ are the thickness and center coordinate of shear layer at each section, respectively. This figure shows self-similarity of velocity data for the FOV located at $x=90$ cm.

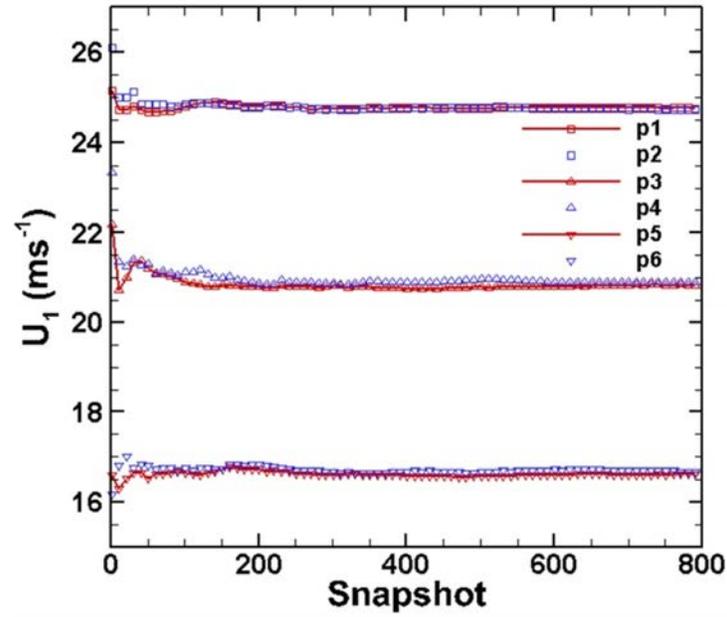

(a)

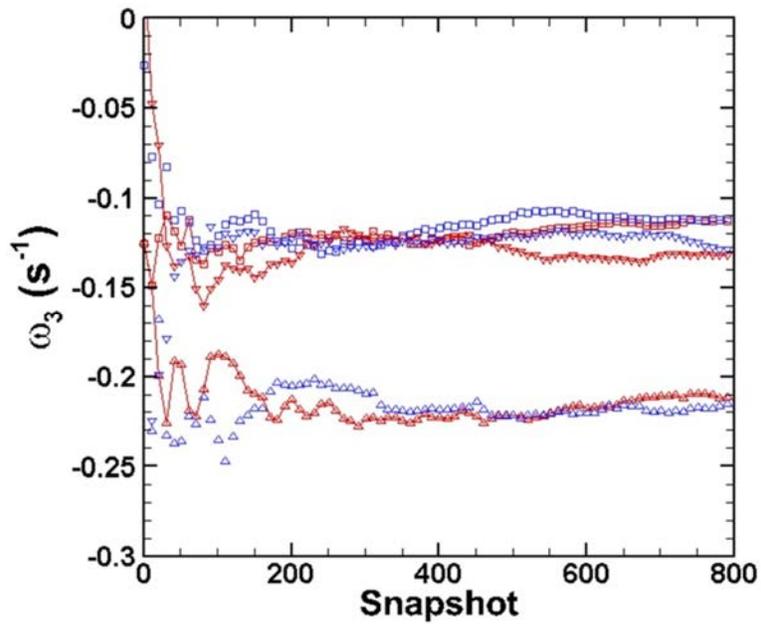

(b)

Figure 5. (a) Cumulative average of stream-wise velocity, (b) Cumulative average of out-of-plane vorticity. The averaging is considered at six points where their location is indicated at Figure 4(a).

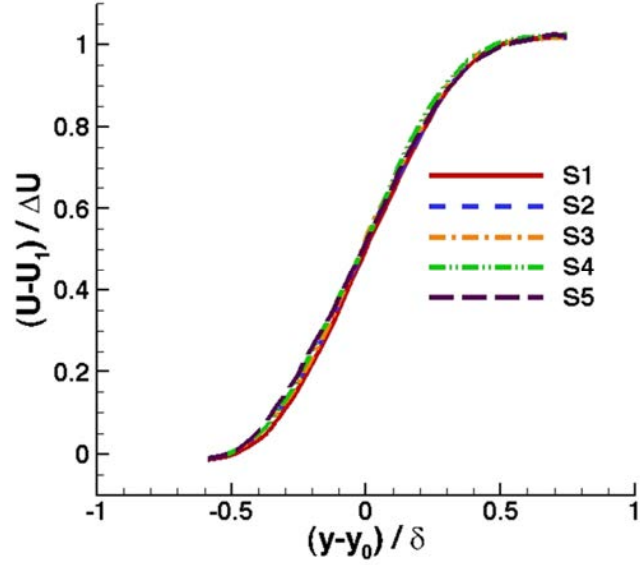

Figure 6. Variations of ensemble average of normalized total velocity with respect to non-dimensional y axis for five deferent sections in x direction (The locations of the sections S1-S5 are illustrated in Figure 3).

Figure 7 compares the normalized root mean square of normal Reynolds stresses calculated using present velocity data with those obtained in previous studies, for a similar velocity ratio [23, 24, 25, 26]. The results show a similar trend where differences are due to non-identical experimental conditions in each case.

## 2.4 Velocity reconstruction in gappy points

The velocity snapshots of SPIV contain 3 percent gappy points on average. About 5% of existing data points is randomly selected as artificial gaps in the present work. The iterative repair procedure of POD defines a mean square error between actual and estimated velocities at spatial locations where real data exist [27]. The defined error provides a basis to calculate a new snapshot eigenfunction by solving a system of linear equations. This new snapshot eigenfunction is used to calculate the new velocity field at gappy points. The repair algorithm is comprehensively presented in the Appendix and the results are examined as follows.

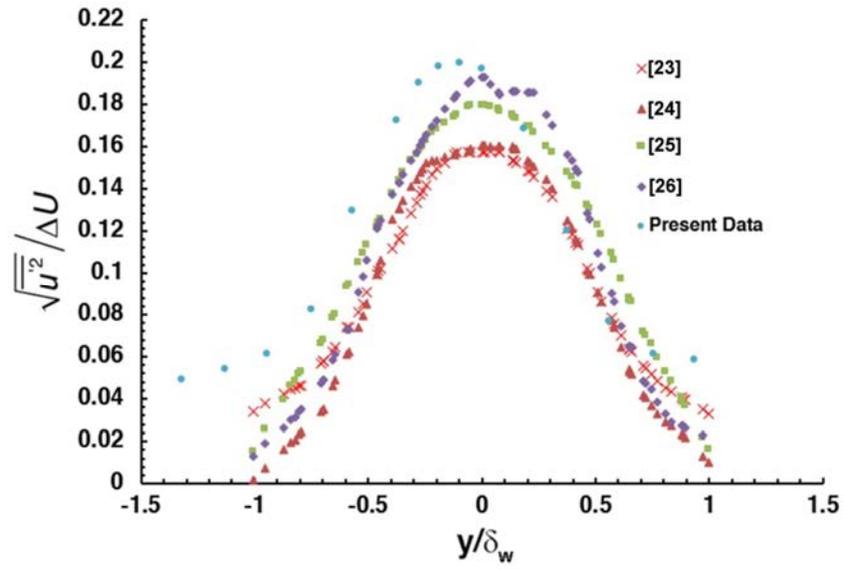

**(a)**

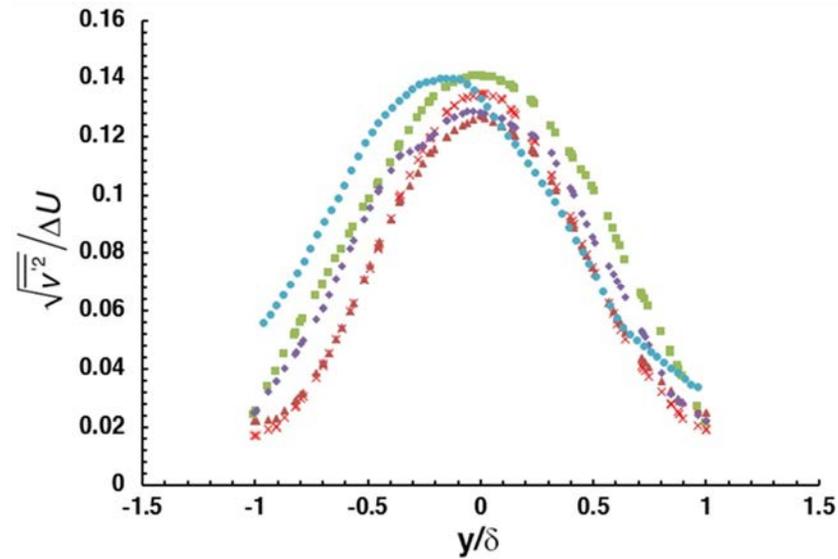

**(b)**

Figure 7. Normalized root mean square of normal Reynolds stress components calculated using present velocity data (center of FOV located at x=90 cm) compared to previous studies [23, 24, 25, 26]: (a) for the x-component, (b) for the y-component, (c) for the z-component. Continued…

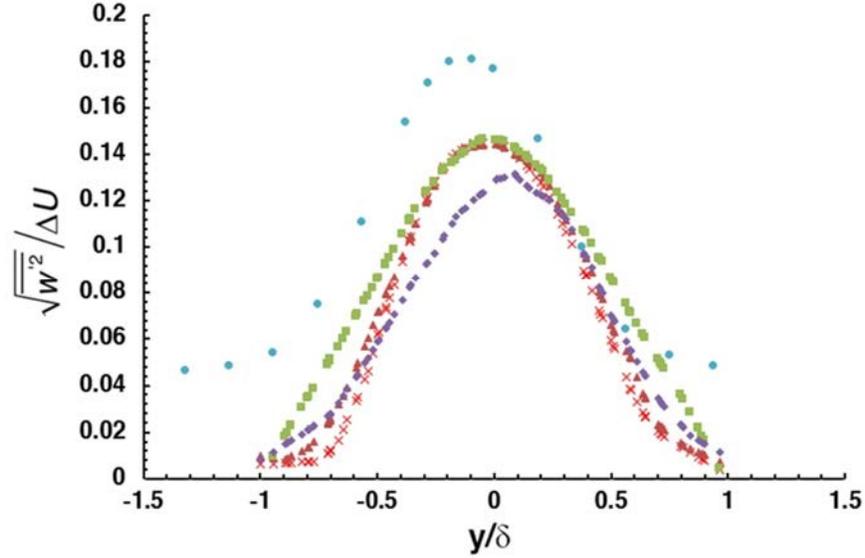

**(c)**

Figure 7. Continued.

*2.4.1 Reconstruction results*

More accurate POD reconstruction results from a larger number of snapshots [28]. In addition, the efficient number of modes must be calculated for each reconstruction of the velocity field, before using the repair algorithm. This process for the present data includes three levels with 200, 400 and 600 snapshots (levels I, II and III, respectively). The efficient number of modes for these levels is calculated to be 58, 115, and 165, respectively.

The POD eigenvalues represent the amount of energy in a given mode [27, 29]. Summation of the eigenvalues, $\lambda^{(s)}$, over all modes gives a measure of the total energy. Figure 8 shows the eigenvalue of each mode for the three above-mentioned levels. Comparison of eigenvalues shows a sudden change of trend in the vicinity of the efficient mode [30]. This behavior is magnified for level I and it repeats for levels II and III. Since the total energy is constant, this trend means that a smaller number of modes for a respective smaller number of snapshots should cover the same energy as levels with more snapshots.

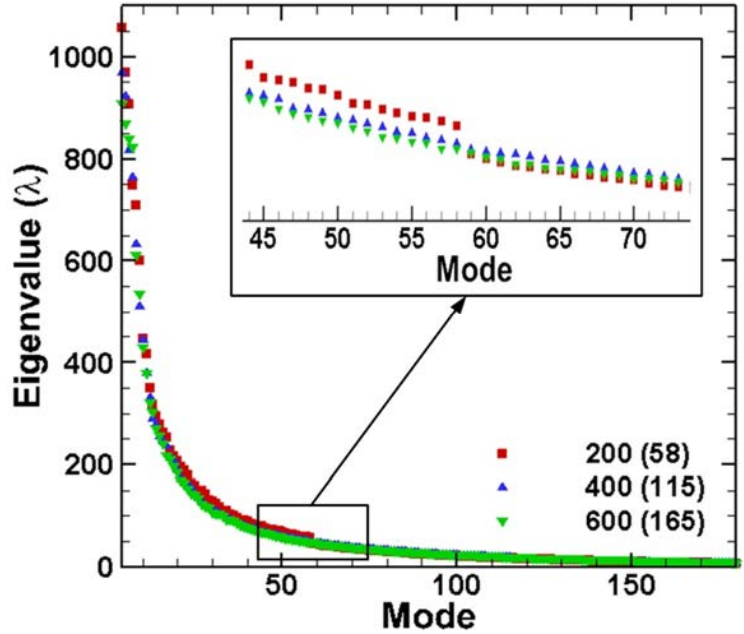

Figure 8. Eigenvalues of reconstructed velocity field for the three levels with respect to corresponding modes. For each level, the efficient number of modes is indicated inside the parenthesis.

Figure 9 presents the error for reconstructions versus the iteration number for levels I, II and III. It shows that suitable convergence is after 5 iterations. The error is smaller for larger number of snapshots. In an efficient reconstruction, it is expected that this error should be less than or close to the measurement uncertainty of the SPIV setup [27]. According to Figure 9, the real reconstruction error for level III (600 snapshots) is about 1.75% which is fairly close to the uncertainty of measured velocity (1.44%). Consequently, based on a compromise between reconstruction computational efficiency and its accuracy with respect to measurement error, the reconstruction with 600 snapshots is selected. The total energy must also converge through succeeding iterations. The convergence of total energy in the present article (not shown here) is after 15 iterations.

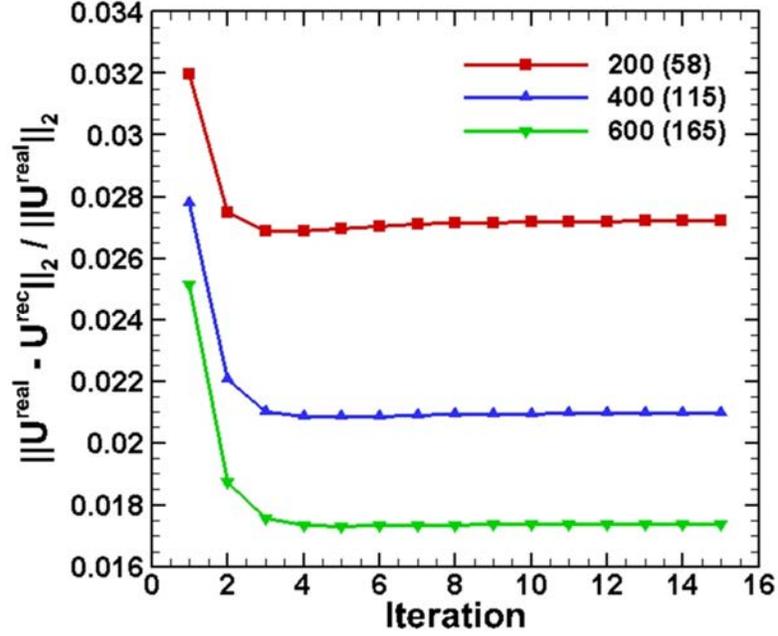

Figure 9. Real error versus iteration number for three reconstruction levels. For each level, the efficient number of modes is indicated inside the parenthesis.

The quality of velocity reconstruction in real gaps is evaluated in Figure 10 which presents the 2D velocity fluctuations (Subtraction of averaged velocity from the instantaneous velocity) in an instantaneous snapshot with high percentage of real gappy points, that are reconstructed by different number of snapshots (levels I, II and III). Gray vectors are the reconstructed velocity fluctuations at real gappy points and the measured velocity fluctuations are in black. The criterion for visual evaluation of velocity vectors is the spatial harmony of reconstructed vectors with their neighboring measured vectors. According to Figure 10, although the reconstructions with 200 and 400 snapshots are consistent in some points, that with 600 snapshots is consistent in almost all the gappy points. The reconstructed velocity vector maps based on 600 snapshots are therefore used in the following sections.

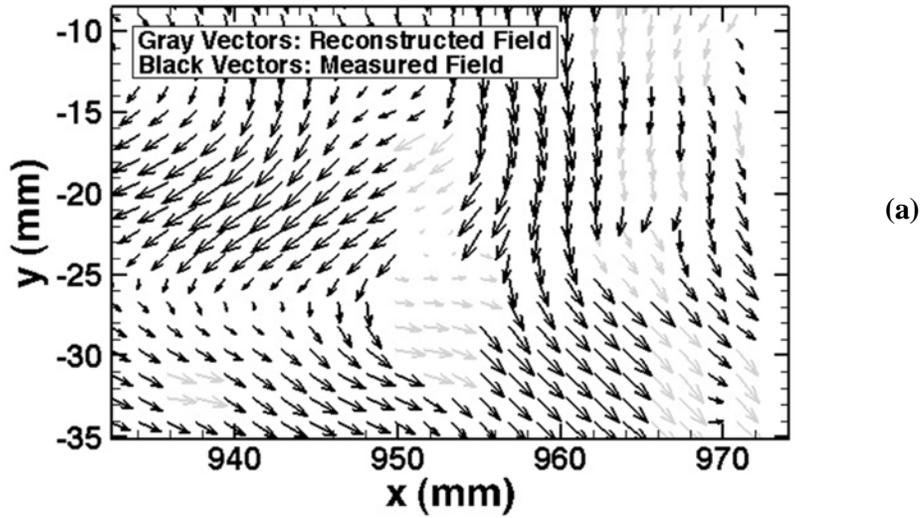

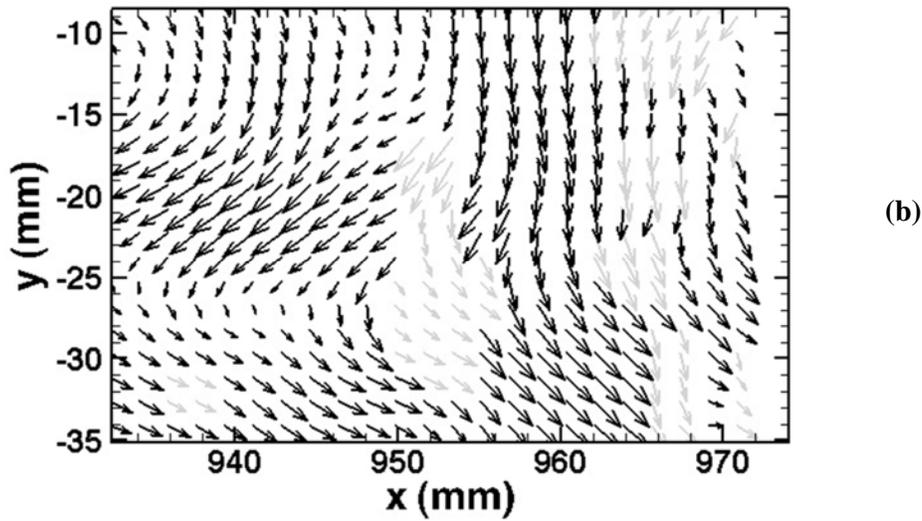

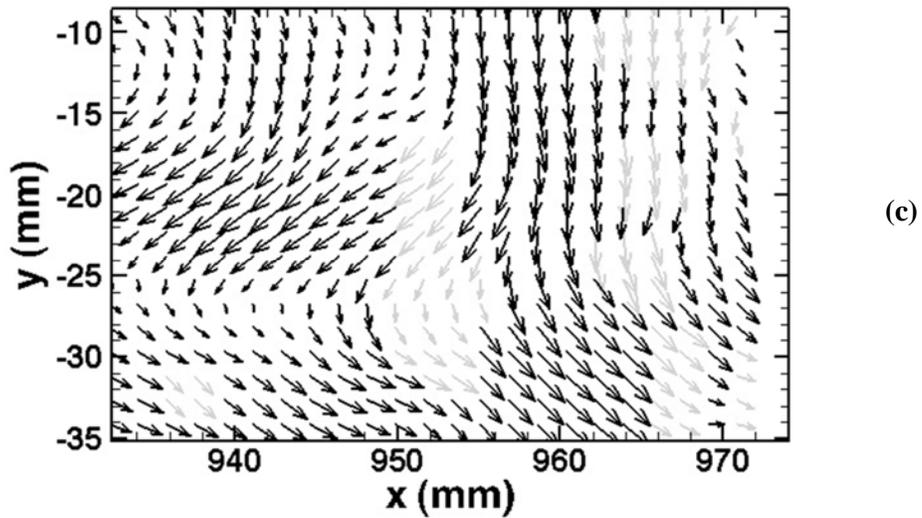

Figure 10. Measured velocity fluctuations at an arbitrary snapshot and the reconstructed velocity fluctuations based on three levels of snapshots: (a) 200 snapshots, (b) 400 snapshots, and (c) 600 snapshots.

### 3. Preparation for a-priori analysis

Herein, the a-priori analysis focuses on the prediction of SGS stress and SGS dissipation parameters. In the previous section, the gappy points within the SPIV data of the turbulent mixing layer flow were reconstructed. In this section, the a-priori analysis of the mixed model SGS predictions before and after removing the high energy modes of the flow is performed.

*3.1 The role of the flow low energy modes*

The purpose of SGS modeling is to apply the effect of SGS scales on the large resolved scales of the flow and solve the closure problem of turbulence. Nevertheless, due to a lack of information about the SGS scales, the parameters of resolved scales such as the filtered velocity ($\overline{U}$) are used to model SGS parameters.

The resolved scales contain several length and time scales with different significance. According to the turbulent energy spectrum concept, the energy containing eddies are in the small wave numbers that correspond to large length scales of the flow. These eddies interact with mean flow and are affected by boundary condition and geometry [31]. In the other sides, small flow scales and small eddies, are proportionally less affected by the boundary condition and geometry. Besides, non-linear interactions between scales are significantly local and it is expected that SGS scales affect neighboring scales. Another important behavior of flow scales is that the scales located in Inertial Sub-Range (ISR) influence each other and are not affected by the scales of energy containing eddies and dissipation range scales [32].

The last paragraph emphasizes that in modeling SGS parameters, it might be advantageous to use small scales of the resolved flow field and especially small scales of filtered velocity field. The multiscale philosophy by Hughes *et al.* [8] and the work by Couplet *et al.* [9], support the idea of present work that is based on removing the high energy modes of the flow in the calculation of the low energy filtered velocity in the next sub-section. Therefore, only small scales of the flow are utilized to model the SGS parameters in the Smagorinsky term of mixed model.

*3.2 Mixed model after removing the high energy modes*

After removing the high energy POD modes of the flow, a low energy filtered velocity ($\overline{U}^P$) is generated to calculate the parameters of the Smagorinsky term within the mixed model in Eq. 5 instead of $\overline{U}$ itself.

$$\overline{U}_i^P(n,m;k) = \sum_{s=k}^{N} a^{(s)}(k) \psi_i^{(s)}(n,m). \qquad k > 1, i = 1,2 \qquad (6)$$

where $\psi$ is the orthogonal POD basis function (or spatial eigenmode), $a$ is the orthonormal temporal coefficient (or orthonormal snapshot eigenfunction) and $s$ is the POD mode number. In addition, $i$ represents any of orthogonal coordinates, $n$ and $m$ are the number of the spatial grid points (in the PIV plane), $k$ represents the number of temporal realizations, $k$ is the smallest (most energetic) used POD mode number and $N$ is the total number of POD modes.

After removing the high energy modes, the SGS stress based on the Smagorinsky model is:

$$\tau_{ij} = -2\nu_t^P \overline{S}_{ij}^P \qquad (7)$$

where $\overline{S}_{ij}^P$ and $\nu_t^P$ are defined as:

$$\overline{S}_{ij}^P = \frac{1}{2}\left(\frac{\partial \overline{U}_i^P}{\partial x_j} + \frac{\partial \overline{U}_j^P}{\partial x_i}\right) \qquad (8a)$$

$$\eta_t^p = (C_{Smag}^p \mathrm{D})^2 \left|\overline{S}^p\right| \tag{8b}$$

where $C_{Smag}^p$ is the Smagorinsky model coefficient after removing the high energy modes, and $\left|\overline{S}^p\right|$ is defined as:

$$\left|\overline{S}^p\right| = \sqrt{2\overline{S_{ij}}^p \overline{S_{ij}}^p} \tag{9}$$

Eventually, the mixed model is postulated as:

$$\tau_{ij} = C_{Sim}\left(\overline{\overline{U_i}\,\overline{U_j}} - \overline{\overline{U_i}}\,\overline{\overline{U_j}}\right) - 2(C_{Smag}^p \mathrm{D})^2 \left|\overline{S}^p\right|\overline{S_{ij}}^p \tag{10}$$

where $C_{Sim}$ and $C_{Smag}^p$ are its unknown coefficients. Table 2 presents the abbreviations for real and modeled SGS quantities in the present article. In addition to the Table 2, the word "After" is used henceforth to show the mixed model prediction after removing high energy modes.

The mixed model has two coefficients: one for the similarity model ($C_{Sim}$) and the other is the Smagorinsky model coefficient ($C_{Smag}$, $C_{Smag}^p$). These coefficients must be set in a way that the mixed model can accurately predicts SGS dissipation [2]. Due to the fact that in mixed models there are two unknown coefficients, only SGS dissipation balance between real and modeled values is not sufficient. To overcome this problem, the balance between real and modeled values of SGS stress is also considered. The fact that the similarity term is more accurate than the Smagorinsky term for prediction of SGS stress, results in considering a balance between real SGS stresses and those modeled by similarity term which gives the similarity-term coefficient in the mixed model:

Table 2. Abbreviations of real and modeled SGS quantities.

| Keyword | Real values | Similarity model | Mixed model | Mixed model (After removing high energy modes of the flow) |
|---|---|---|---|---|
| Abbreviation | R | S | M | MA |

$$C_{ij}^{Sim} = \frac{\langle t_{ij}^{Real} \rangle}{\langle \overline{\overline{\overline{U_i U_j}}} - \overline{\overline{\overline{U_i}}}\,\overline{\overline{\overline{U_j}}} \rangle}, \text{ No summation on i, j} \quad (11)$$

where $t_{ij}^{Real}$ is the real SGS stress component, $C_{ij}^{Sim}$ is its corresponding similarity-term coefficient and the symbol $\langle \cdot \rangle$ represents the averaging operation. Eq. (11) shows that the similarity-term coefficient of mixed models is individually calculable for each SGS stress component.

Now, the only unknown coefficient is the coefficient of eddy-viscosity term, i.e. the Smagorinsky coefficient. A balance between SGS dissipation modeled by the eddy-viscosity term of the mixed model and sufficient share of real SGS dissipation (that is the share not predicted by the similarity term) calculates this coefficient. The SGS energy dissipation estimated by similarity term of mixed model is subtracted from the real SGS dissipation (the nominator of Eqs. (12) and (13)), in order to evaluate the underestimation of real SGS dissipation by the similarity-term. The eddy-viscosity term will compensate such underestimation. The Smagorinsky coefficients based on the above balance are calculated as:

$$C_{Smag} = \sqrt{\frac{\langle t_{ij}^{Real} \overline{S_{ij}} - t_{ij}^{Sim} \overline{S_{ij}} \rangle}{\langle -2D^2 |\overline{S}| \overline{S_{ij}}\, \overline{S_{ij}} \rangle}} \quad (12)$$

$$C_{Smag}^{P} = \sqrt{\frac{\langle t_{ij}^{Real}\overline{S_{ij}} - t_{ij}^{Sim}\overline{S_{ij}}\rangle}{\langle -2\Delta^{2}|\overline{S}^{P}|\overline{S_{ij}}^{P}\overline{S_{ij}}\rangle}} \qquad (13)$$

where $t_{ij}^{Sim}$ is the component of SGS stress modeled by similarity term of the mixed model which is defined as:

$$t_{ij}^{Sim} = C_{ij}^{Sim}\left(\overline{\overline{U_i U_j}} - \overline{\overline{U_i}}\,\overline{\overline{U_j}}\right), \text{ No summation on i, j} \qquad (14)$$

where $C_{ij}^{Sim}$ is calculable by Eq. (11).

## *3.3 Calculation of gradient-based quantities*

The planar SPIV data for the mixing layer in the present study provides all three velocity components as a basis for the aforementioned a-priori analysis of SGS models. Velocity component gradients normal to measurement plane (z direction) are unavailable for single-plane measurement. However, it is possible to calculate the gradient of the out-of-plane velocity component normal to the measurement plane ($\frac{\partial \overline{U_3}}{\partial x_3}$) for incompressible flow based on the continuity equation:

$$\overline{S_{33}} = \frac{\partial \overline{U_3}}{\partial x_3} = -(\overline{S_{11}} + \overline{S_{22}}) \qquad (15)$$

where 1, 2 and 3 denotes *x*, *y* and *z* directions, respectively.

Out of plane shear strains ($\overline{S_{13}}$ and $\overline{S_{23}}$) are not available and calculation of the quantity $\overline{S_{ij}}\,\overline{S_{ij}}$ and the SGS energy dissipation needs assumptions. Two approaches are reported in the literature. First, isotropy is assumed hence the same value is considered for all cross terms. Although this assumption is not accurate for anisotropic flows, it was used in studies dealing with anisotropic turbulent flows such as flows with rapid straining stages [33], turbulent jet flow [3] and flow through rotor and stator blades of a centrifugal turbomachine [34]. Second, the two-dimensional surrogate of the quantity $\overline{S_{ij}}\,\overline{S_{ij}}$ and SGS energy dissipation from the existing components of strain and stress tensors is considered [35, 36, 37, 38, 39]. Although this approach does not include all the contributing components, the limiting states of turbulence are not assumed and the calculation is based on the available data. The latter approach is used in the present study, to estimate the quantity $\overline{S_{ij}}\,\overline{S_{ij}}$ and SGS energy dissipation ($P$):

$$\overline{S_{ij}}\,\overline{S_{ij}} \approx \left( \langle \overline{S_{11}}\,\overline{S_{11}} \rangle + \langle \overline{S_{22}}\,\overline{S_{22}} \rangle + \langle \overline{S_{33}}\,\overline{S_{33}} \rangle + 2\langle \overline{S_{12}}\,\overline{S_{12}} \rangle \right) \qquad (16)$$

$$P \approx - \left( \langle \tau_{11}\overline{S_{11}} \rangle + \langle \tau_{22}\overline{S_{22}} \rangle + \langle \tau_{33}\overline{S_{33}} \rangle + 2\langle \tau_{12}\overline{S_{12}} \rangle \right) \qquad (17)$$

### 3.4 Filter kernel

The filter kernel used in present study is the top hat filter [1]:

$$G(r) = \prod_{i=1}^{3} H\left( \frac{\Delta_{(i)}}{2} - |r_{(i)}| \right) \qquad (18)$$

where $H$ is the Heaviside function and $\Delta_{(i)}$ represents the filter width in each direction. The top-hat filter is a common filter type [12] where the averaging nature of the top-hat filter in a spatial domain, provides stable implementation of this filter kernel in a numerical code in the physical space.

There are two filtering operations in scale similarity model as well as in the similarity term of mixed model. The size of first filter for all the filtering operations is considered to be $\Delta_1 = 3d_T$ ($d_T$ is defined as $d_T = \sqrt{d_x d_y}$, where $d_x$ and $d_y$ are PIV grid spacing in x and y directions, respectively). Two different sizes are considered for the second filter, namely $D_2 = 7d_T$ and $D_2 = 5d_T$ which are represented as case 1 and case 2, respectively.

## 4. Results and discussion

In the present study, low energy POD modes of the flow are used to calculate a low energy filtered velocity ($\overline{U}^P$) and the corresponding strain rate components in the Smagorinsky term of the mixed model (See Eqs. (6-10)). Figure 11 evaluates contribution of different velocity modes (the first three and the 100$^{th}$ POD modes) on the overall velocity for the x-component ($U_1$). The first velocity mode generates a velocity field which is very similar to the mean flow and contains considerably higher values of velocity in comparison with other modes. The first POD mode also contains about 99 percent of total kinetic energy. Other POD modes show a different velocity field and share a small contribution of total kinetic energy. Comparison of the 100$^{th}$ POD velocity mode with the second and third modes shows that the higher velocity modes represent smaller flow structures. The results were also similar for other velocity components.

In order to evaluate the role of low energy flow structures on the SGS parameter prediction, the first POD mode, is removed and the low energy filtered velocity (See Eq. (6)) is then calculated. The similarity and mixed model estimation of SGS stress/dissipation, before and after removing the first mode, is evaluated based on a-priori analysis that is presented in the next sections for two different filter size. The calculated model coefficients are first reported. Then a-priori prediction of the SGS dissipation/stress by the similarity and mixed model is compared with real ones.

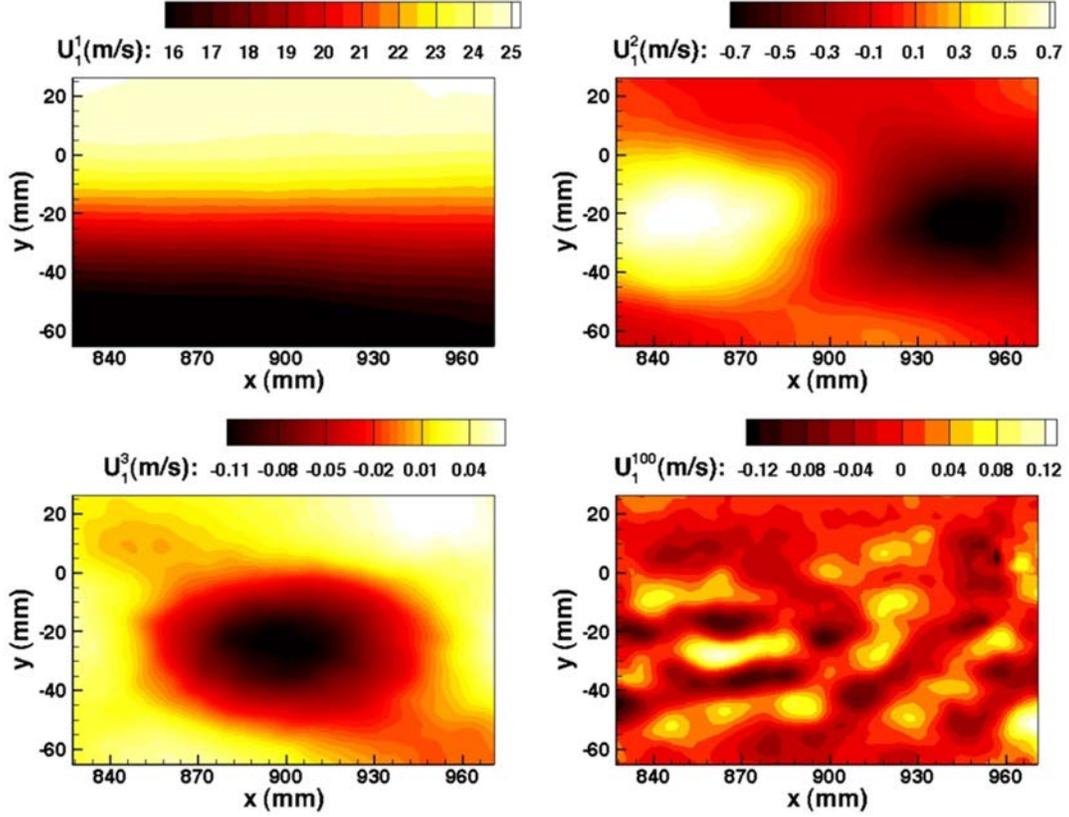

Figure 11. The first three and the 100$^{th}$ POD velocity modes for the velocity component in x- direction ($U_1$). In legend of each contour plot, the superscript of U represents the mode number.

*4.1 Strain rates*

Contours of $\overline{S_{12}}^p$ (reduced strain rate, See Eq. (8a)) and $\overline{S_{12}}$ (actual strain rate) for an instantaneous velocity snapshot are presented in Figure 12. The difference between reduced and actual strain rates in this figure is due to absence of mean flow velocity gradients in the calculation of reduced strain rate. The spatial balance between positive and negative values of $\overline{S_{12}}^p$ is stronger in comparison with those of $\overline{S_{12}}$.

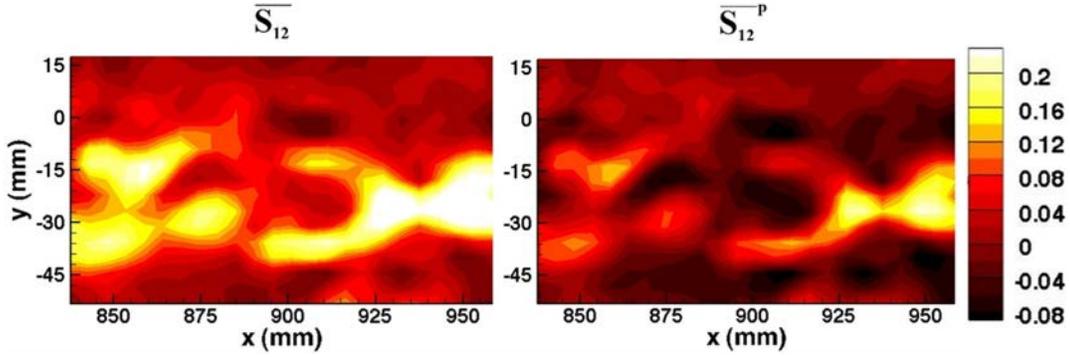

Figure 12. Contours of $\overline{S_{12}}$ and $\overline{S_{12}}^p$ for an instantaneous snapshot.

In present study, normal and shear stresses are assessed from $t_{11}$ and $t_{12}$, respectively. Although the latter is the only available shear stress component based on a single-plane SPIV measurement, it is the dominant component of shear stress and strongly affects SGS quantities for the current SPIV setup and mixing layer.

## *4.2 Model Coefficients*

Table 3 presents the calculated similarity and mixed model coefficients for cases 1 and 2 in comparison with the calculated coefficients for various SGS models in the previous studies. The Smagorinsky coefficient is an order of magnitude smaller than the similarity coefficients, whether calculating it separately or within the mixed model [2, 14, 33]. The similarity coefficient from previous studies for mixed or similarity model varies from 0.11 to 1.75. This wide range of values is attributed to flow specifications, filter type and size and flow Reynolds number [14, 33, 40]. A-priori analysis based on the balance of real and modeled SGS dissipation for a jet/wake flow field showed that calculated similarity coefficient could be in the range of 0.11 to 0.38 [14].

Table 3. Comparison of the calculated coefficients for various SGS models based on the present and previous studies.

| Reference | Flow type | Velocity field extraction method | Investigated SGS models | A-priori calculated model coefficients |
|---|---|---|---|---|
| Present study | Turbulent mixing layer flow | SPIV | Similarity, Mixed model (Similarity plus Smagorinsky), | **Case 1** $C_{11}^{Sim}$=0.2227, $C_{12}^{Sim}$=0.2187, $C_{Smag}$=0.0233, $C_{Smag}^{p}$=0.0292 **Case 2** $C_{11}^{Sim}$=0.41, $C_{12}^{Sim}$=0.3995, $C_{Smag}$=0.0131, $C_{Smag}^{p}$=0.0164 |
| Liu et al. [2] | Far field of a turbulent round jet | 2C-PIV | Smagorinsky, Similarity, Mixed model | $C_{Smag}$=0.08-0.1, $C_{Sim}$=0.6-0.9, $C_{Smag}^{Mixed}$=0.07-0.1, $C_{Sim}^{Mixed}$=1.0-1.1 |
| Liu et al. [33] | Turbulence with rapid straining | 2C-PIV | Smagorinsky, Similarity | Before straining: $C_{Smag}$=0.06-0.14, $C_{Sim}$=1.0 During constant straining: $C_{Smag}$=0.14, $C_{Sim}$=0.45 |
| Chen et al. [35] | Turbulence subjected to rapid straining and destraining | 2C-PIV | Smagorinsky, Mixed model | $C_{Smag}$=0.1-0.21, $C_{Smag}^{Mixed}$=0.09-0.19 ($C_{Sim}$ wasn't reported) |
| Cook [40] | Turbulent jet | Experiment | Similarity | $C_{Sim}$=0.54-1.75 |
| Akbari and Montazerin [14] | Flow field in the rotor exit region of a centrifugal turbomachine | SPIV | Smagorinsky, Similarity | $C_{Smag}$=0.0434-0.0641 $C_{Sim}$=0.1124-0.3877 |

The similarity coefficient was assumed unity based on the assumption that SGS and Leonard stresses are of the same order of magnitude [3]. The calculated similarity coefficients in the present study are smaller than unity for cases 1 and 2. A smaller size of second filter in case 2 results in larger similarity coefficients. This is attributed to smaller captured Leonard stress for the smaller second filter (in case 2), since the Leonard stress captures the scales between grid (first) and test (second) filters [32].

The Smagorinsky coefficients are calculated in the present study such that underestimation of SGS dissipation by similarity term is compensated. The second filter length is larger in case 1 than case 2, which makes the underestimation and the required compensation of SGS dissipation stronger, leading to larger values for Smagorinsky coefficients.

*4.3 SGS dissipation prediction*

Table 4 presents the total average of real and modeled SGS energy dissipation. The total average is calculated after averaging the data in all snapshots and all spatial SPIV grid points. The Smagorinsky coefficient of the mixed model (before and after removing the first mode) is calculated such that the mixed model exactly predicts the total average of real SGS dissipation. The Smagorinsky term in the mixed model fills the gap between the real SGS dissipation and those under-predicted by similarity term. The total averaged SGS dissipation from similarity model shows a larger estimation error for case 1 than that of case 2. A smaller second filter in case 2 results in more accurate estimation of SGS dissipation.

Figure 13 shows the SGS dissipations averaged in homogeneous directions in order to effectively assess the model performance across the mixing layer. The ensemble average for case 1, is again averaged in the stream-wise direction (x direction) and variation of the averaged SGS dissipation in the y direction is shown. The symbols represent the total average of SGS energy dissipation (over all snapshots and entire spatial domain). The similarity model underestimates SGS dissipation, especially in the mixing region (approximately in the middle of y-axis). Considering the purpose of introduction of mixed model, which is provision of an appropriate prediction for SGS dissipation, the mixed model must predict the SGS dissipation correctly before and after removing the first mode. In Figure 13, this purpose is achieved before and after removing the first mode. However, the differences between the mixed model predictions are expected for other turbulence quantities such as SGS stresses.

Table 4. Total average of real and modeled SGS energy dissipation.

|  | Total average of SGS energy dissipation ($m^2 s^{-3}$) | | | |
|---|---|---|---|---|
|  | Real value | Similarity model prediction | Mixed model prediction | Mixed model prediction (After) |
| Case 1 | 0.0072 | 0.0068 (Error: 5.55 %) | 0.0072 | 0.0072 |
| Case 2 | 0.0072 | 0.0070 (Error: 2.77 %) | 0.0072 | 0.0072 |

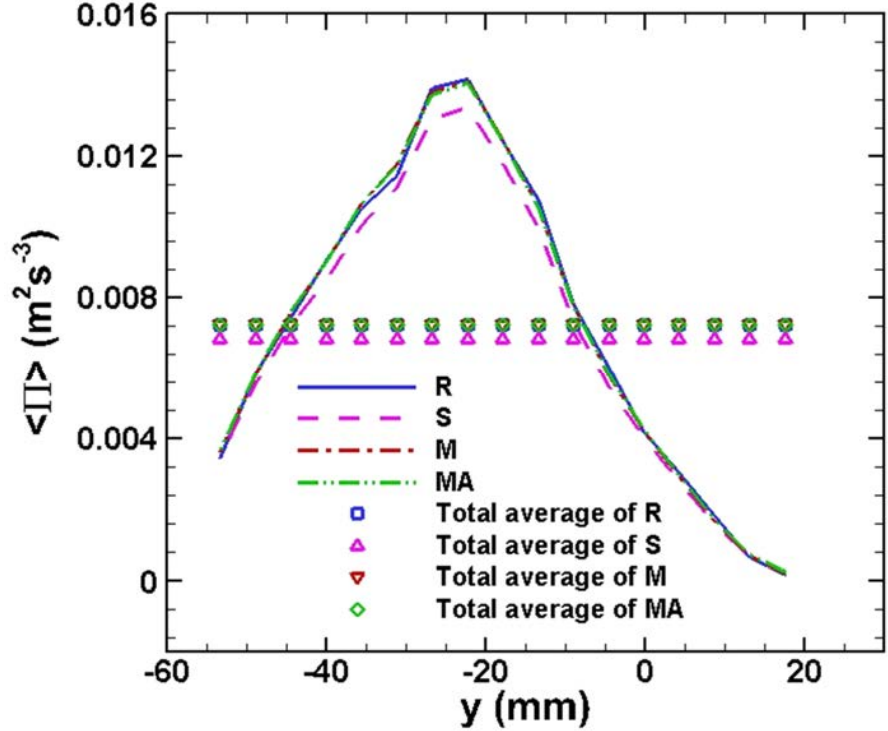

Figure 13. The ensemble average of SGS dissipation for real and modeled values, after averaging in the x-direction. The symbols represent the total average of SGS dissipation over all snapshots and the entire spatial domain.

Figure 14 presents the probability density function (PDF) of prediction error for ensemble averaged SGS dissipation in case 1, for similarity and mixed model, as compared with that of real SGS dissipation. Prediction error is the difference between real and modeled SGS dissipations divided by the real SGS dissipation. The mixed model predictions, before and after removing the first mode, are almost similar. They are more accurate in comparison with similarity model prediction. Similar graphs to those of Figures 13 and 14 were generated for case 2, which repeated similar patterns and showed the smaller extent of underestimation of SGS dissipation by the similarity model in comparison with case 1.

The correlation coefficient between real and modeled SGS dissipation ($P^{Real}$ and $P^{Mod}$), as a scalar evaluation tool is calculated as:

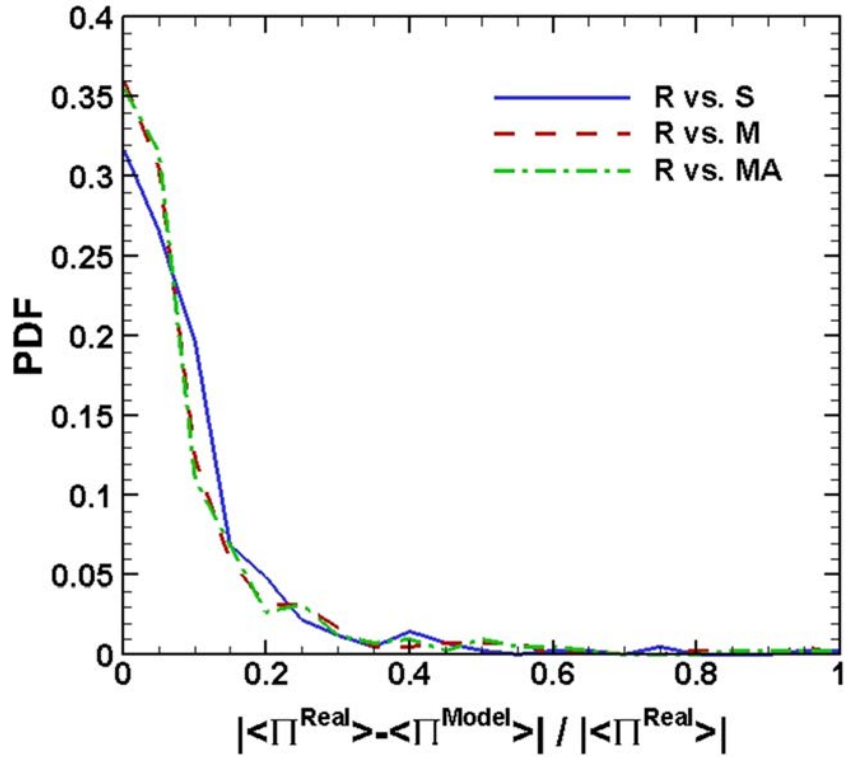

Figure 14. PDF of prediction error for ensemble averaged SGS dissipation.

$$r(P^{Mod}, P^{Real}) = \frac{\langle P^{Mod} P^{Real}\rangle - \langle P^{Mod}\rangle\langle P^{Real}\rangle}{\left[\left(\langle (P^{Mod})^2\rangle - \langle P^{Mod}\rangle^2\right)\left(\langle (P^{Real})^2\rangle - \langle P^{Real}\rangle^2\right)\right]^{1/2}} \quad (19)$$

A correlation coefficient equal to unity represents an exact prediction where multiplication of the modeled term by a scalar coefficient results in the real one. The correlation coefficient based on SGS dissipations is presented for similarity and mixed model predictions in Table 5. It is found that adding the eddy-viscosity term to similarity term in mixed model does not affect the correlation coefficient. Case 2, where the ratio of second filter to first filter size was 5/3, showed stronger correlated SGS dissipation predictions as compared with case 1, where the ratio of second to first filter size was 7/3.

*4.4 SGS stress prediction*

The similarity model prediction is assumed to be the real value of total averaged SGS stress because SGS stress was balanced to calculate the similarity coefficient. The added Smagorinsky term in the mixed model, also balances the real and modeled SGS dissipations.

Tables 6 and 7 present the relative error of total averaged SGS stress ($\tau_{11}$ and $\tau_{12}$) predicted by mixed model before and after removing the first mode. Before removing the first mode, the prediction error of mixed model is about ten times larger than when the first mode is removed. Additionally, the error of total average of $\tau_{12}$ is notably larger than that of $\tau_{11}$, which might be due to the dominancy of shear stress in shear flows. Comparison of the calculated data for cases 1 and 2 shows that the model error is reduced for smaller second filter length, which is consistent with the trend for SGS dissipation.

Table 5 - Correlation coefficient between real and modeled SGS dissipation.

|        | Similarity model | Mixed model | Mixed model (After) |
|--------|------------------|-------------|---------------------|
| Case 1 | 0.93             | 0.93        | 0.93                |
| Case 2 | 0.96             | 0.96        | 0.96                |

Table 6. Total average of $t_{11}$ for real and modeled stresses.

|              | Total average of $t_{11}$ ($m^2s^{-2}$) | | | |
|--------------|------------|-----------------------------|--------------------------------|----------------------------------|
| Real & Model | Real value | Similarity model prediction | Mixed model prediction         | Mixed model prediction (After)   |
| Case 1       | 0.088366   | 0.088366                    | 0.088377 (Error: 0.012 %)      | 0.088365 (Error: 0.001 %)        |
| Case 2       | 0.088366   | 0.088366                    | 0.088369 (Error: 0.003 %)      | 0.088366 (Error: ~ 0 %)          |

Table 7. Total average of $t_{12}$ for real and modeled stresses.

|              | Total average of $t_{12}$ ($m^2s^{-2}$) | | | |
|--------------|------------|-----------------------------|--------------------------------|----------------------------------|
| Real & Model | Real value | Similarity model prediction | Mixed model prediction         | Mixed model prediction (After)   |
| Case 1       | -0.01419   | -0.01419                    | -0.01478 (Error: 4.15 %)       | -0.01426 (Error: 0.49 %)         |
| Case 2       | -0.01419   | -0.01419                    | -0.01438 (Error: 1.33 %)       | -0.01421 (Error: 0.14 %)         |

Figure 15 presents ensemble average of SGS stress ($t_{11}$ and $t_{12}$), which is averaged in x-direction, for case 1 over y direction. Part (a) shows almost similar predictions of similarity and mixed model for $t_{11}$. Before removing the first mode, the deviation of $t_{12}$ predicted by mixed model from the similarity model prediction and real values, is larger than when the first mode is removed. The deviation is generally larger in the mixing region due to more intensive turbulent interactions in this region. The general trends for prediction of similarity and mixed model for case 2 are similar with those of case 1, except for the extent of deviations which is smaller for case 2.

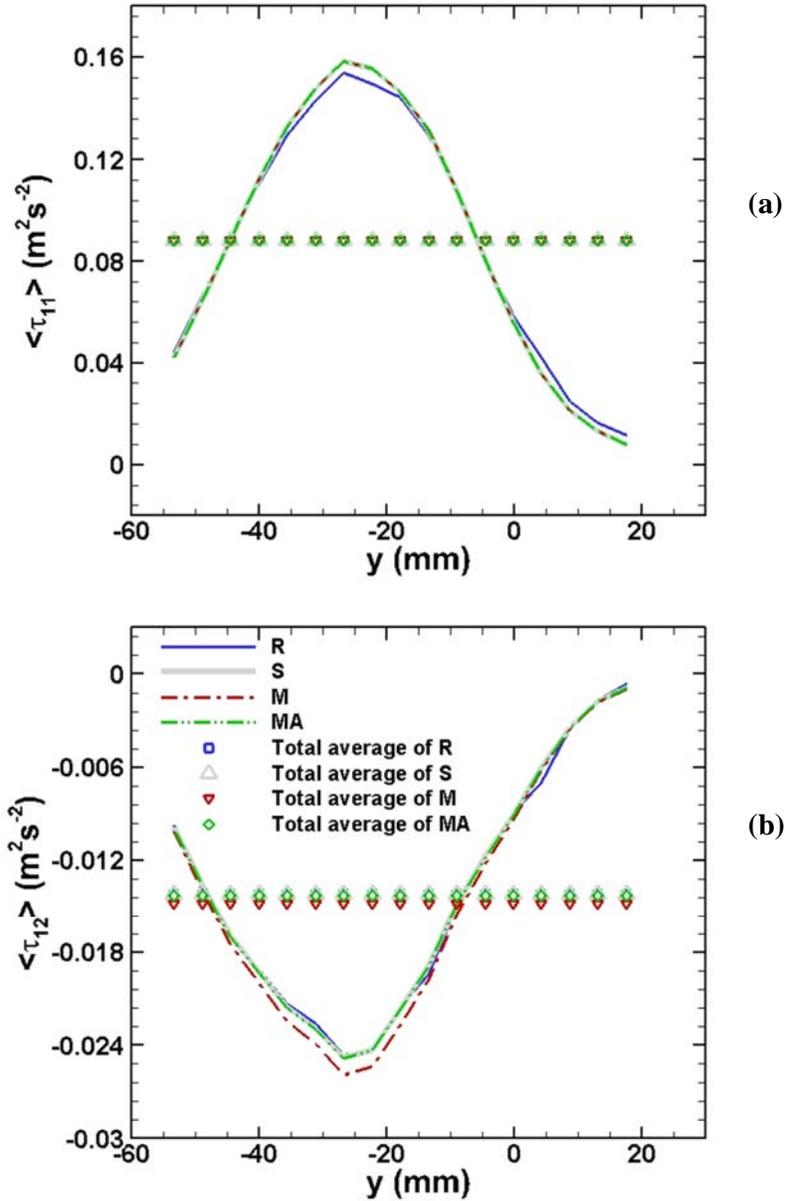

Figure 15. Variation of the ensemble averages of modeled and real SGS stresses for case 1 in the y-direction. The ensemble averages are also averaged in x-direction; (a) for $t_{11}$, and (b) for $t_{12}$.

A detailed comparison of real SGS stress and those modeled by similarity and mixed model is presented in Figure 16 where variations of the ensemble average of $t_{12}$ in y direction is plotted at six sections in the x-direction. Because of minor differences between model predictions for $t_{11}$, this figure and the next ones only deal with the dominant stress component, i.e. $t_{12}$. Figure 16 confirms that the predictions of similarity and mixed model, after removing the first mode, are coincident and closer to the real SGS stress in comparison with the condition before removing the first mode. The largest deviations between model predictions and the real values occur around the central region of the mixing layer with similarly larger values of shear SGS stress.

Figures 15 and 16 show that SGS stress distribution is asymmetric with respect to $y=0$ which is in the same level as the separator plate. Additionally the center of mixing region, which is attributed by the maximum absolute value of shear SGS stress, is not aligned with $y=0$ and is located below this coordinate. This downward tendency of the mixing region is an inherent characteristic of a classical confined shear layer that is due to entrainment of high speed stream into the low speed stream [1, 41].

Figure 17 compares the instantaneous predictions of mixed model with those of similarity model at each spatial point, before and after removing the first mode. The predicted values of $t_{12}$ at different points of each snapshot are considered as a 2D matrix. The difference between similarity-based matrix and each of the mixed model-based matrices and their calculated 2-norms, is plotted in Figure 17 to evaluate SGS stress predictions of mixed model before and after removing the first mode. The average of all plotted data is also presented in the figure. The deviation of $t_{12}$ between mixed model and similarity model after removing the first mode, is less than the deviation before removing the first mode.

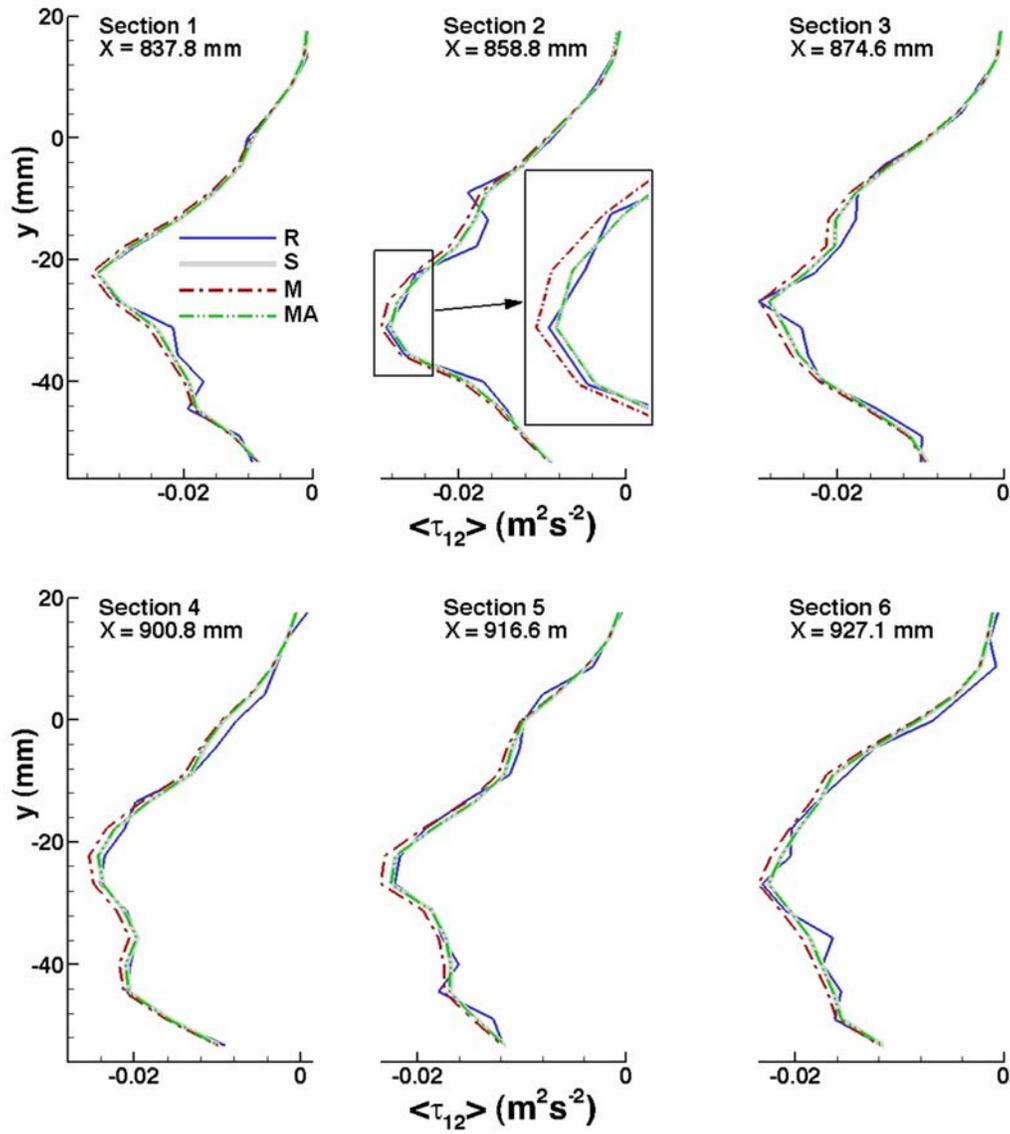

Figure 16. Variations of ensemble average of $t_{12}$ for real and modeled values in y direction at six sections spaced in the x-direction. The insert in section 2 is depicted to show model predictions and the coincidence of similarity and mixed model values after removing the first mode.

The results of this study were already presented for $k = 2$ (See Eq. (6)), in which the first POD mode is removed and the second and higher POD modes are used in the calculations. However, in order to analyze effect of the parameter $k$ on the SGS prediction, Figure 17 is also drawn for other values of $k$ starting from 3 to N, which N is the total number of POD modes. In such cases, the number of removed modes is more than one (from 2 to N-1), respectively. The average of the deviation of instantaneous $t_{12}$ predicted by the mixed model (after removing $k-1$ POD modes) from the similarity model prediction, which were shown in Figure 17 with dashed line, is presented in Figure 18. Herein, the solid horizontal line corresponds to the original mixed model prediction ($k=1$) without removing any POD mode. Figure 18 shows that the most accurate prediction for instantaneous values of $t_{12}$ is for $k=2$, that corresponds to removing just the first POD mode, as is the case in the presented results throughout the article.

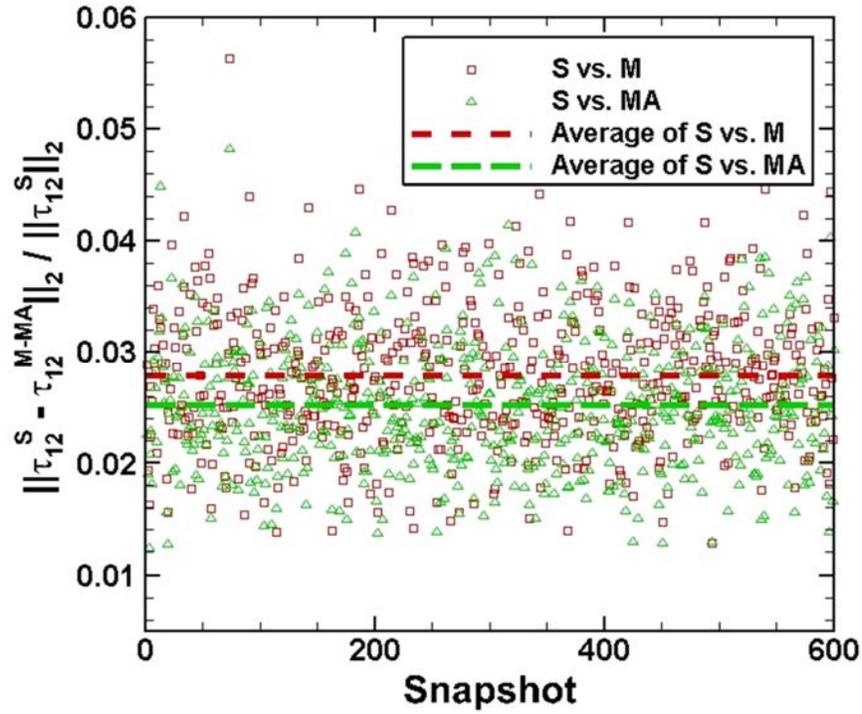

Figure 17. Deviation of instantaneous $t_{12}$ predicted by mixed model (before and after removing the first mode) from similarity model prediction based on the 2-norm.

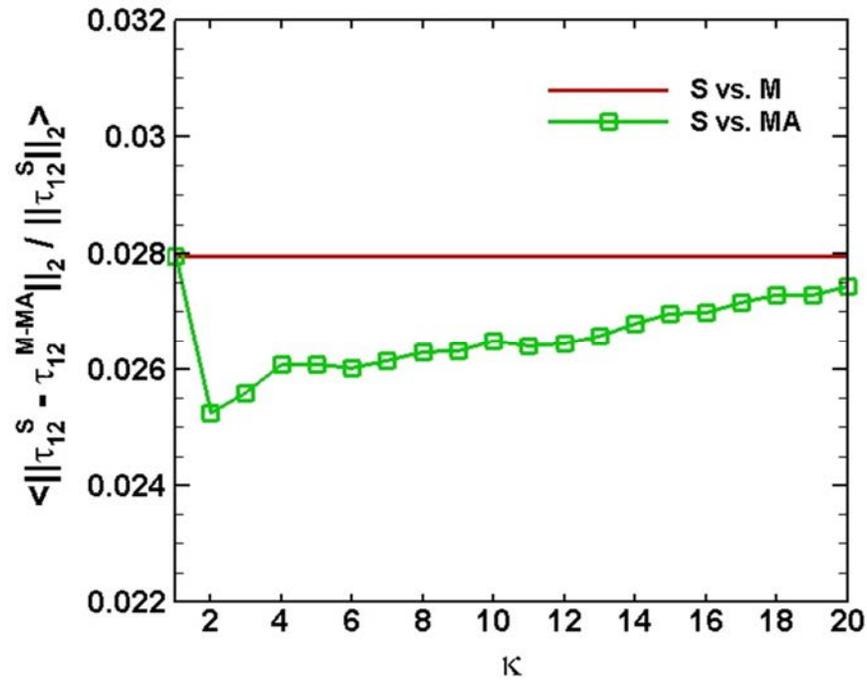

Figure 18. Average of the deviation of instantaneous $t_{12}$ predicted by the mixed model (after removing $k-1$ POD modes) from the similarity model prediction based on 2-norm.

Figure 19 presents PDF of relative prediction error for ensemble average of $t_{12}$ for similarity and mixed models. The relative prediction error is defined as the ratio of real and modeled SGS stress difference to the real SGS stress component. The PDFs of similarity and mixed model, after removing the first mode, are close and indicate better model performance in comparison with the condition before removing the first mode.

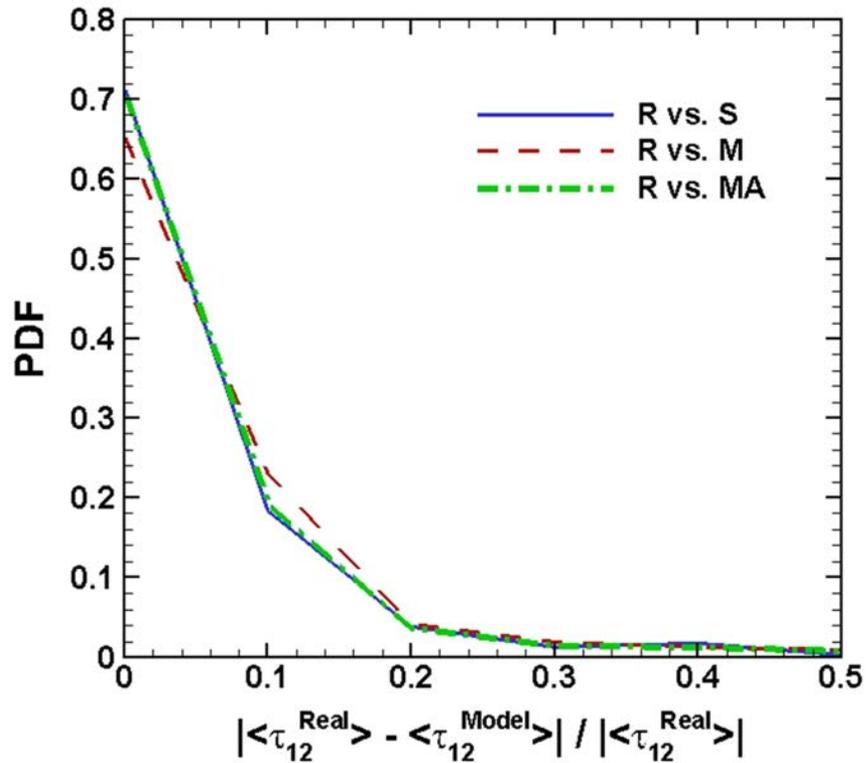

Figure 19. PDF of relative prediction error for ensemble average of $t_{12}$.

## 5. Conclusion

In the present study, the effects of low energy POD modes of the flow on the SGS parameter prediction made by the mixed model were evaluated. In order to conduct such analysis, the a-priori approach using the experimental velocity data of a turbulent mixing layer flow from SPIV measurements was selected. Due to existence of missing or erroneous data in the SPIV measurements, first, the velocity vectors were reconstructed using the GPOD method. Three levels of velocity snapshots were used and the most accurate reconstruction was chosen to conduct the a-priori analysis.

In order to evaluate the effects of low energy modes on SGS parameter prediction, the strain rates in the Smagorinsky term of the mixed model were calculated after removing the high energy modes. In this way, the small flow structures (that have more universal characteristics) are used in the calculation of strain rates. In present work, removing only the first POD mode (most energetic mode) was more advantageous in SGS parameter prediction. The a-priori analysis showed that after removing the first mode, the mixed model properly predicted the SGS dissipation and also the SGS stresses were predicted more accurately in comparison with the condition before removing the first mode. With removing the first mode and using the other low energy modes in the calculation of the strain rates, a different spatial-temporal distribution for SGS stress was predicted by the mixed model such that the SGS dissipation remained correct and moreover, the average of SGS stress was no more affected by the deviations enforced from the eddy-viscosity term.

**Appendix: SPIV data reconstruction algorithm**

Gappy proper orthogonal decomposition, which is a form of proper orthogonal decomposition method, is widely used to reconstruct the missing data extracted by PIV [27, 30, 42, 43, 44]. POD is a linear procedure for extracting a basis for modal decomposition from an ensemble of signals and dimension reduction in complex non-linear problems [45, 46]. POD modes are sorted based on their contribution to the total energy in a way that the first mode contains the largest percentage and the last mode contains the smallest percentage of the total energy. Although all data modes are calculated in this method, only a finite number of modes contain prime data features.

Sirovich [47] introduced the snapshot POD in 1987, to solve the eigenvalue problem created in POD process. The orthogonal eigenfunctions are considered as linear combinations of the velocity at different snapshots:

$$\phi_i^{(s)}(n,m) = \sum_k a^{(s)}(k) U_i(n,m;k) \qquad (A1)$$

where $\phi$ and $a$ are spatial eigenmodes and snapshot eigenfunctions, respectively. In addition, $i$ represents component of the velocity and eigenmode, $s$ is the mode number, $k$ is the snapshot number indicator and $n$ and $m$ show spatial discretized points.

The intermediate eigenvalue problem is given as:

$$\sum_k C_{k,l} a^{(s)}(k) = \lambda^{(s)} a^{(s)}(l) \qquad (A2)$$

where $\lambda$ is the eigenvalue and $l$ is the snapshot number indicator. The kernel $C_{k,l}$ is

expressed as:

$$C_{k,l} = \frac{1}{K} \sum_n \sum_m U_i(n,m;k) U_i(n,m;l) \tag{A3}$$

where $K$ is the total number of used snapshots.

Solving Eq. (A2), provides the snapshot eigenfunctions, $a^{(s)}(k)$ and then using Eq. (A1), the spatial eigenmodes, $f_i^{(s)}(n,m)$, are calculable. The velocity field can be reconstructed then, using the orthonormality of the snapshot eigenfunctions.

$$U_i(n,m;k) = \sum_s a^{(s)}(k) f_i^{(s)}(n,m) \tag{A4}$$

Gunes *et al.* [28] introduced a method to find the efficient number of POD modes, $q$, for data reconstruction. In this method, some of the valid existing data points are considered as gappy points (artificial gaps), and after the reconstruction, the real error (the difference between real and estimated data at artificial gaps) is defined as:

$$\frac{\|U_M^{(i)} - U\|_2}{\|U\|_2} = \frac{\sqrt{\int\int_{\Omega T}(U_M^{(i)}(x,t) - U(x,t))^2 \, dx \, dt}}{\sqrt{\int\int_{\Omega T}(U(x,t))^2 \, dx \, dt}} \tag{A5}$$

where $U$ is the actual total velocity at artificial gaps which is equal to:

$$U = \sqrt{U_1^2 + U_2^2 + U_3^2} \tag{A6}$$

In addition, $U_M^{(i)}$ is the corresponding reconstructed total velocity in the $i^{\text{th}}$ iteration using $M$ modes. Besides, $\Omega$ and $T$ represent the spatial and temporal domains, respectively. The real error can be defined for velocity components, as well. The number of POD modes which provides the minimum value for real error, is the efficient number of modes and adding the modes in repair procedure is continued until achievement to such minimum.

*Data repair algorithm*

The iterative repair procedure introduced by Everson and Sirovich [42] is used to reconstruct the gappy points in velocity snapshots. The location of gappy points is initially addressed by a defined matrix $m$:

$$m(n,m;k) = \begin{cases} 1 & \text{if } U_i(n,m;k) \text{ exists} \\ 0 & \text{if } U_i(n,m;k) \text{ is missing or erroneous} \end{cases} \quad (A7)$$

The ensemble average values are used as initial guess to fill the gappy points:

$$U_i(n,m;k) = \begin{cases} U_i(n,m;k) & \text{if } m(n,m;k) = 1 \\ \langle U_i \rangle(n,m) & \text{if } m(n,m;k) = 0 \end{cases} \quad (A8)$$

The iterative procedure is as follows:

(1) The snapshot POD is used to calculate the spatial eigenmodes, $f_i^{(s)}(n,m)$, using equations (A1) - (A3).

(2) An estimate of the velocity field, $\tilde{U}_i(n,m;k)$, is defined using an unknown series of orthonormal eigenfunctions, $b^{(s)}(k)$, with $q$ terms:

$$\tilde{U}_i(n,m;k) = \sum_{s=1}^{q} b^{(s)}(k) f_i^{(s)}(n,m) \quad (A9)$$

(3) A mean square error, $\tilde{E}(k)$, is introduced between actual and estimated velocities at spatial locations where real data exists. The mask $m(n,m;k)$ ensures that only points with real data are added in the summation:

$$\tilde{E}(k) = \sum_n \sum_m E(n,m;k)\mathfrak{m}(n,m;k) \qquad (A10)$$

where

$$E(n,m;k) = \left(\tilde{U}_i(n,m;k) - U_i(n,m;k)\right)^2 \qquad (A11)$$

(4) A minimum mean square error can be calculated after differentiating Eq. (A10) with respect to unknown coefficients, $b^{(s)}(k)$, and equating the result to zero. Consequently, a system of linear equations with $K$ equations and unknowns results that its solution provides the unknown coefficients, $b^{(s)}(k)$:

$$[M]_{s',s}[B]_{k,s} = [V]_{k,s'} \qquad (A12)$$

where

$$\begin{aligned} [M]_{s',s} &= \sum_n \sum_m f_i^{(s)}(n,m) f_i^{(s')}(n,m)\mathfrak{m}(n,m;k) \\ [B]_{k,s} &= b^{(s)}(k) \\ [V]_{k,s'} &= \sum_n \sum_m U_i(n,m;k) f_i^{(s')}(n,m)\mathfrak{m}(n,m;k) \end{aligned} \qquad (A13)$$

(5) Now, the obtained coefficients, $b^{(s)}(k)$, are used in Eq. (A9) to estimate the velocity field.

(6) The missing data is replaced at gappy points with the velocity estimates obtained in the previous step.

$$U_i(n,m;k) = \begin{cases} U_i(n,m;k) & \text{if } \mathfrak{m}(n,m;k) = 1 \\ \tilde{U}_i(n,m;k) & \text{if } \mathfrak{m}(n,m;k) = 0 \end{cases} \qquad (A14)$$

(7) Return to step one, if convergence is not achieved.

Murray and Ukeiley [27], calculated the difference between the estimated coefficients, $b^{(s)}(k)$, and those calculated directly from POD, $a^{(s)}(k)$. The repair iteration stops when the coefficients, $b^{(s)}(k)$ and $a^{(s)}(k)$, are closer than a predefined threshold. The convergence error is defined as:

$$e_{conv} = \frac{\sqrt{\sum_k \sum_s \left(a^{(s)}(k) - b^{(s)}(k)\right)^2}}{\sqrt{\sum_k \sum_s \left(a^{(s)}(k)\right)^2}} \quad (A15)$$

In the present study, the repair iteration stops when $e_{conv} < 0.0001$.

**Acknowledgement**

The authors are sincerely grateful to Dr. Zohreh Mansoori and Energy Research Institute of Amirkabir University of Technology.